\documentclass[12pt]{article} 
\renewcommand\appendix{\par
  \setcounter{section}{0}
  \setcounter{subsection}{0}
  \setcounter{figure}{0}
  \setcounter{table}{0}
  \renewcommand\thesection{Appendix \Alph{section}}
  \renewcommand\thefigure{\Alph{section}\arabic{figure}}
  \renewcommand\thetable{\Alph{section}\arabic{table}}
}

\usepackage{amsmath}  
\usepackage{amsfonts} 
\usepackage{graphicx} 
\usepackage[T1]{fontenc}
\usepackage{bm}
\usepackage{appendix}
\usepackage{amssymb}

\def\theequation{\arabic{section}.\arabic{equation}}

\newcommand{\be}{ \begin{equation}}
\newcommand{\ee}{\end{equation}} 
\begin{document} 
\def\theequation{\arabic{section}.\arabic{equation}} 
\begin{titlepage} 
\title{Analogues of glacial 
valley profiles in particle mechanics and in cosmology} 
\author{Valerio Faraoni$^1$ and 
Adriana M. Cardini$^2$\\ \\ 
{\small \it $^1$~Physics Department and Environmental 
Science 
Programme, Bishop's University}\\ 
{\small \it 2600 College Street, Sherbrooke, Qu\'ebec Canada 
J1M~1Z7}\\
{\small vfaraoni@ubishops.ca}\\
{\small \it $^2$~Physics Department, Bishop's University}\\ 
{\small \it 2600 College Street, Sherbrooke, Qu\'ebec 
Canada  J1M~1Z7}\\
{\small acardini15@ubishops.ca}
}
\date{} \maketitle 
\thispagestyle{empty} 
\vspace*{1truecm} 
\begin{abstract} 
An ordinary differential equation describing the transverse 
profiles of U-shaped glacial valleys, derived with a 
variational principle, has two formal analogies which we 
analyze. First, an analogy with point particle mechanics 
completes the description of the solutions. Second, an 
analogy with the Friedmann equation of relativistic 
cosmology shows that the analogue of a glacial valley 
profile is a universe with a future singularity but 
respecting the weak energy condition. The equation  
unveils also a Big Freeze 
singularity, which was not observed 
before for positive curvature index.
\end{abstract} 

\begin{center} 
{\bf Keywords:} glacial valley transverse profiles, 
glaciology-cosmology analogies, glaciology-mechanics 
analogies, Friedmann equation. 
\end{center} 
\end{titlepage}


\section{Introduction}
\setcounter{equation}{0}
\label{sec:1}

It has long been acknowledged in glaciology since its inception 
\cite{Campbell, McGee}, and it is common knowledge also in 
elementary geography, that valleys carved by glaciers are U-shaped  
while valleys carved by the action of rivers are V-shaped. 
Here we focus on the former. 
The detailed and continued process of reshaping a valley by a glacier via 
erosion of the valley walls and bed over time is not  
simple and is best modelled with numerical techniques  
\cite{numerical, numerical2, numerical3}. If one 
is interested only in the final result of the glacier 
action, simpler analytic approaches can be used. Given the 
scarcity of analytic models in the literature, theoretical 
approaches to this 
problem are valuable. A clever idea proposed by Hirano 
and Aniya in 1988 \cite{HiranoAniya88} consists of formulating a 
variational principle which extremizes the friction of the 
ice against the valley walls, subject to an appropriate constraint. Let 
the cross-sectional profile of a glacial valley be 
described by a function $y(x)$, where $x$ is a coordinate 
transverse to the glacier flow. Hirano and Aniya argued 
that friction (a functional of the cross-profile $y(x)$) 
should be minimum at the end of the erosion process, 
subject to the constraint that the contact length of the 
cross-profile of the ice is constant. This contact length 
between two endpoints $x_1$ and $x_2$ of the transverse 
profile is \be \label{1} s\left[ y(x) \right] 
=\int_{x_1}^{x_2} ds= \int_{x_1}^{x_2} \sqrt{dx^2+dy^2} = 
\int_{x_1}^{x_2} \sqrt{ 1+(y')^2}\, dx =\mbox{const.} , \ee 
where a prime denotes differentiation with respect to $x$. 
The friction force is modelled by Coulomb's law as $ f=\mu 
N $, where $\mu$ is the 
friction coefficient and the normal force is $ N=\rho g h 
A_1$. Here $\rho $ is the ice density, $g$ is the 
acceleration of gravity, $h$ is the ice thickness, and $A_1$ 
is the area of contact between the ice and the 
bed.\footnote{If present, water pressure between the 
glacier and its bed is treated as constant and does not 
contribute to the variational principle \cite{HiranoAniya88}.}

By considering a unit width of ice in the longitudinal 
direction 
of the glacier, the friction force due to an element 
of contact length $ds$ is $df=\mu P ds$. 
Further, $P=\eta (y_s-y)$, where  $y_s$ denotes the ice 
surface and $\eta$ is a constant. Extremizing the friction 
\be
f\left[ y(x) \right] =\mu\eta \int_{x_1}^{x_2} (y_s-y) \sqrt{ 1+(y')^2} \, dx
\ee
subject to the constraint~(\ref{1}) leads to 
\be
\delta J = \mu\eta \, \delta \int_{x_1}^{x_2} \left( y_s - 
y+\lambda \right) \sqrt{ 1+ (y')^2} \, dx \equiv \delta 
\int_{x_1}^{x_2} L=0 \,,
\ee
where $\lambda$ is a Lagrange multiplier and $L$ is the Lagrangian. The 
Euler-Lagrange equation
\be\label{EulerLagrange}
\frac{d}{dx} \left( \frac{ \partial L}{\partial y'}\right) 
-\frac{ \partial L}{\partial y}=0 
\ee
yields the ODE \cite{HiranoAniya88}
\be \label{6}
\frac{ y_s - y+\lambda}{\sqrt{ 1+(y')^2}} =c_1 \,,
\ee
where $c_1$ is a constant. This equation is also  
obtained by solving the classic catenary problem of  
mechanics ({\em e.g.}, \cite{Goldstein}, 
see Appendix~\ref{appendix:A}  
for further comments) and, therefore, it is not surprising that
 Hirano and 
Aniya obtained catenary solutions of eq.~(\ref{6}) 
\cite{HiranoAniya88}. 

Hirano and Aniya's method and conclusions 
\cite{HiranoAniya88} have been criticized by Harbor 
\cite{Harbor90} (see also the subsequent debate 
\cite{HiranoAniya90, Morgan05, HiranoAniya05}). First, the 
assumptions in the model of \cite{HiranoAniya88} are 
inconsistent with other common assumptions in glaciology 
\cite{Harbor90}. Second, friction should be {\em 
maximized}, not minimized \cite{Harbor90} 
(although physically important, 
this change does not affect the first order variational 
principle, which only requires the friction integral to be 
{\em extremized}).  In a reply to Harbor 
\cite{HiranoAniya90}, Hirano and Aniya agree on this 
point but stand by the validity of application of the 
variational principle and of their previous result. Further 
critique by Morgan appeared fifteen years later 
\cite{Morgan05}; he pointed out that there is no 
physical basis for the isoperimetric constraint~(\ref{1}), 
which 
should be replaced by the requirement that 
the area of the cross-section of the glacial valley be kept 
fixed instead. The rationale is that, by considering a unit 
width of ice in the direction of longitudinal flow, the ice 
volume is thus kept constant \cite{Harbor90}. Indeed, it 
appears that Hirano and Aniya themselves had originally 
considered such a constraint, as they state in their reply 
to Morgan \cite{HiranoAniya05}, although it did not appear 
in their original paper \cite{HiranoAniya90}. 

The new Lagrangian constraint of Morgan in the variational 
principle leads \cite{Morgan05} to the ODE 
\be \label{7}
\left( \frac{ y'}{y} \right)^2 = \frac{1}{\left( \lambda y-C 
\right)^2}- \frac{1}{y^2} \,,
\ee
where $y(x)$ is now the ice thickness at transverse 
coordinate\footnote{The valley profile is now 
$z(x)=y_s-y(x)$, cf.~fig.~1 of \cite{Morgan05}.}  
$x$, $\lambda$ is again a Lagrange multiplier, and $C$ is a 
constant, with $\lambda >1$ and $C>0$ required in order to 
have a smooth symmetric solution $y(x)$ on the interval 
$\left[ -x_0, x_0 \right]$ with $y'(0)=0$ \cite{Morgan05}. 
Eq.~(\ref{7}) will be adopted in the rest of this 
work to describe transverse profiles of glacial valleys. 
An exact solution of eq.~(\ref{7}) was also provided in 
Ref.~\cite{Morgan05},
\be\label{8}
\left( \lambda^2-1 \right) |x| =C\lambda \sqrt{1-w^2} +C 
\arccos w\,,
\ee
where
\begin{eqnarray}
&& w = \left( \frac{ \lambda^2 -1}{C} \right) y - \lambda 
\,, 
\label{9}\\
&& \nonumber\\
&& -\frac{1}{\lambda} \leq w \leq 1 \,, \;\;\;\;\;\;
\frac{C}{\lambda} \leq y \leq \frac{C}{\lambda-1} 
\,.\label{10}
\end{eqnarray}
Another formal solution of eq.~(\ref{7}) for $\lambda^2 <1$ 
is given in  the recent reference \cite{Gibbons0}:
\be \label{Gibbonssolution}
\pm \frac{\left( 1-\lambda^2 \right)^{3/2}}{|D|} \, x= 
\lambda \sqrt{ w^2-1} +\ln \left| w+\sqrt{w^2-1} \right| +D 
\,,
\ee
where $D$ is an integration constant and $|w|>1$. For $C=0$ 
and 
$|\lambda |<1$ the solutions are linear.

The constraint of fixed cross-sectional area of the valley 
may perhaps seem questionable but no better one  
has been proposed in the literature thus far. The 
requirement of fixed 
cross-sectional area is also used in the numerical 
modelling of the erosion process leading to U-shaped 
valleys \cite{numerical2}. In any case, some 
constraint must be imposed because, if the friction 
integral is maximized without constraints, the first order 
variation 
\be
\delta \int_{x_1}^{x_2} \left( y_s-y \right)\sqrt{ 
1+(y')^2} \equiv  \delta \int_{x_1}^{x_2}L_0 =0 \,,
\ee
where the Lagrangian is now
\be
L_0\left( y(x), y'(x) \right)= \left( y_s-y \right) 
\sqrt{1+(y')^2} \,, \label{12}
\ee
produces an equation which admits only solutions 
which are unphysical. 
In fact, since the Lagrangian~(\ref{12}) 
does not depend explicitly on $x$, the corresponding 
Hamiltonian
\be
{\cal H}_0= p_yy' - L_0
\ee
is conserved, where
\be
p_y = \frac{ \partial L_0}{\partial (y')}= \frac{ \left( 
y_s-y 
\right)y'}{\sqrt{ 1+(y')^2} } 
\ee
is the momentum canonically conjugated to $y$. The 
Euler-Lagrange equation~(\ref{EulerLagrange}) for $L_0$ has 
the 
first integral 
\be \label{xab}
\frac{ \left( y_s-y\right) (y')^2}{\sqrt{ 1+(y')^2}} - 
\left( y_s-y\right)  \sqrt{ 1+(y')^2}=C \,,
\ee
where $C$ is a constant. Using the variable $\xi \equiv 
y_s-y >0 $, eq.~(\ref{xab}) is equivalent to 
\be \label{15}
C \sqrt{1+(\xi ')^2}+\xi=0 \,,
\ee
which requires that $C<0$, hence we set $C \equiv -C_2 $, 
where $C_2>0$ is a constant with the dimensions of a 
length. Since $\xi \geq 0$, one obtains
\be \label{1.17}
\xi' = \pm \sqrt{ \frac{\xi^2}{C_2^2}-1} 
\ee
which requires $\xi \geq C_2>0$. All the solutions 
$\xi(x)$ of 
eq.~(\ref{1.17}) are not bounded from above 
and are given by
\be\label{wrongeq}
y_s-y(x)= C_2^2 \, \mbox{e}^{ \mp \frac{ (x-x_0)}{C_2} }
+\frac{ \mbox{e}^{\pm \frac{(x-x_0)}{C_2} }   }{4}
\ee
(with $x_0$ another integration constant) and require $|x| 
\geq C_2$, which does not describe a valley geometry. 
Therefore, some Lagrangian constraint must be imposed when 
extremizing the friction integral $f[y(x)]$.

The fact that eq.~(\ref{7}) has an analogue in point 
particle mechanics seems to have been missed in the 
glaciology literature, while the fact that it has an 
analogue in the Friedmann equation of cosmology was noted 
in passing in the recent references~\cite{Gibbons0, Gibbons}. As we 
discuss in detail in 
Sec.~\ref{sec:3} below, eq.~(\ref{7})  is a special case of 
Friedmann type equations which are of fundamental importance
 in cosmology. A 
mathematical peculiarity of this type of equations (and 
therefore also of eq.~(\ref{7})) demonstrated in 
\cite{Gibbons} is that 
the graphs of all the solutions (in our case, of the 
transverse valley profiles $y(x)$) 
are {\em roulettes}. A roulette is the locus of a point 
which lies on, or inside, a curve which rolls without 
slipping along a straight line.\footnote{A more general 
definition is that the curve rolls without slipping along 
another curve but, for Friedmann-type equations, the latter 
is taken to be a straight line \cite{Gibbons}.}

Our goal is to explore the analogues of the ODE~(\ref{7}) 
in point particle mechanics and in cosmology, obtaining 
insight into the properties of this equation. In turn, we 
uncover a type of cosmological singularity which was 
studied recently \cite{Mariam} for spatially flat universes 
in the now abundant literature on cosmological 
singularities (\cite{Wald, BarrowGallowayTipler, Caldwell, 
BarrowSFS, Sahni, Dabrowski2, Dabrowski, Fernandez, Diego} 
and references therein).

\section{Particle mechanics analogues of glacial  
valley cross-profiles}
\setcounter{equation}{0}
\label{sec:2}

We follow Ref.~\cite{Morgan05} and we assume that $C>0$ 
(but see the discussion below).  The ODE~(\ref{7}) can be 
rewritten as 
\be\label{2:1}
\frac{(y')^2}{2} +V(y) =E \,, 
\ee
where $(y')^2/2 $ can be regarded as the kinetic energy of  
a particle of unit mass in one-dimensional motion if $x$ 
and $y(x)$ are  
seen as the analogues of time and of the 
one-dimensional position, respectively, while 
\be
V(y) \equiv \frac{-y^2}{2 \left( \lambda y -C \right)^2} 
\ee
is an effective potential energy, and $E=-1/2$ is the 
total mechanical energy of the particle, which is fixed to this 
particular value. Newton's second 
law of motion $ 
y''= -dV/dy $ then rules the  one-dimensional motion of the 
particle subject 
to the conservative force $-dV/dy$, and eq.~(\ref{2:1}) is 
a first integral corresponding to conservation of energy 
$E$ (with the fixed value $E=-1/2$). The solutions 
$y(x)$ corresponding to the possible motions of the 
particle are candidates for the description of  
cross-profiles of glacial valleys. As 
is well known from mechanics, a qualitative understanding 
of the motion can be obtained from the study of the potential 
$V(y)$ and of the intersections between its graph and the horizontal line 
$E=-1/2 $ (which is called ``Weierstrass approach'' in 
various applications  \cite{Weierstrass}-\cite{Weierstrass3}).

\begin{figure}
\centering
\includegraphics[scale=0.65]{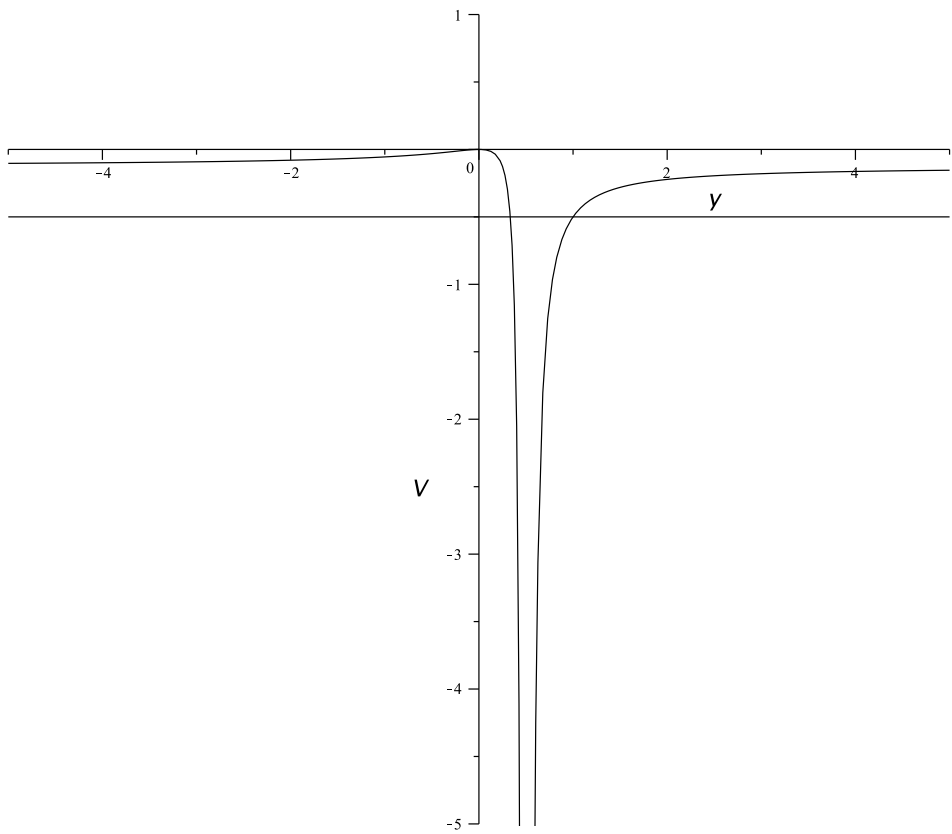}
\caption{The potential $V(y)$ intersecting the 
horizontal line of constant energy $E=-1/2$ (the 
parameter values $\lambda=2$, $C=1$ are chosen for 
illustration).  \label{fig:V}}
\end{figure}

It is  $V(y)<0 $ ~$\forall y\neq 
0$, ~$V(0)=0$, ~$V(y) \rightarrow -1/(2\lambda^2) $ as $y 
\rightarrow \pm \infty$, there is a vertical asymptote 
$y=C/\lambda >0$, and 
\be
V'(y)= \frac{Cy}{ \left( \lambda y- C \right)^3} \,,
\ee
therefore the function $V(y)$ is increasing for $y<0$ and 
for $y>C/\lambda$, decreasing for $0< y < C/\lambda$, and 
maximum at $y=0$ (see fig.~\ref{fig:V}). 

We look for regions of bounded motions $y(x)>0$, 
corresponding to finite ice thickness. We restrict to the 
situation $C>0, \lambda >0$, for which the vertical 
asymptote $y=C/\lambda$ of $V(y)$ lies in the $y>0$ region. 
However, eq.~(\ref{7}) and the potential $V(y)$ are 
invariant under the exchange $\left( C, \lambda \right) 
\rightarrow \left( -C, -\lambda \right)$ and, formally, the 
situation $C<0, \lambda <0$ is the same as the one that we 
describe below.

\subsection{The case $\lambda>1$}

If $E=-\frac{1}{2} < -\frac{1}{2\lambda^2}$, corresponding 
to $| \lambda |>1 $ (and we take $\lambda>1$ here), the 
horizontal line $E=-1/2$ describing the conserved 
energy of the particle associated with
the glacial valley cross-profile lies below the 
horizontal 
asymptote of $V(y)$ (fig.~\ref{fig:V}). There are two 
regions corresponding to bounded motions $y(x) >0$ (we 
ignore the region $y<0$ because it is meaningless for the 
glacial valley problem). The first region is
\be\label{interval1}
0 < y_1 \leq y(x) < \frac{C}{\lambda} \,, 
\ee 
while the second region is 
\be\label{interval2}
\frac{C}{\lambda} < y(x) \leq y_2 \,,
\ee
where $y_{1,2}$ are turning points. The condition $\lambda 
>1$ is the condition for bounded solutions stated in 
Ref.~\cite{Morgan05}, which now receives a graphical 
interpretation. 

The particle cannot attain the position $y=C/\lambda$ where 
the potential becomes singular and is therefore confined to 
either the region~(\ref{interval1}) or the 
region~(\ref{interval2}). 

The turning points $y_{1,2}$ are found analitically by 
setting equal to zero the kinetic energy $(y')^2/2$ of the 
particle\footnote{This condition corresponds to zero slope 
of the valley profile $y(x)$, therefore to its lowest 
point.} which leads, using eq.~(\ref{7}), to the quadratic 
algebric equation
\be
\left( \lambda^2-1 \right)y^2 -2\lambda Cy +C^2=0 \,.
\ee
The roots are 
\be \label{y12}
y_{1,2}= \frac{C}{\lambda \pm 1} \,.
\ee
The range 
\be
 \frac{C}{\lambda} <y(x) \leq \frac{C}{\lambda -1}
\ee
reproduces the condition~(\ref{10})
reported in Ref.~\cite{Morgan05}, while the
range
\be
\frac{C}{\lambda +1} \leq y(x) < \frac{C}{\lambda}
\ee
 does not appear in the analysis of this reference, which 
is therefore augmented by the graphical mechanical analogy.

\subsection{The case $\lambda=1$}

\begin{figure}
\centering
\includegraphics[scale=0.65]{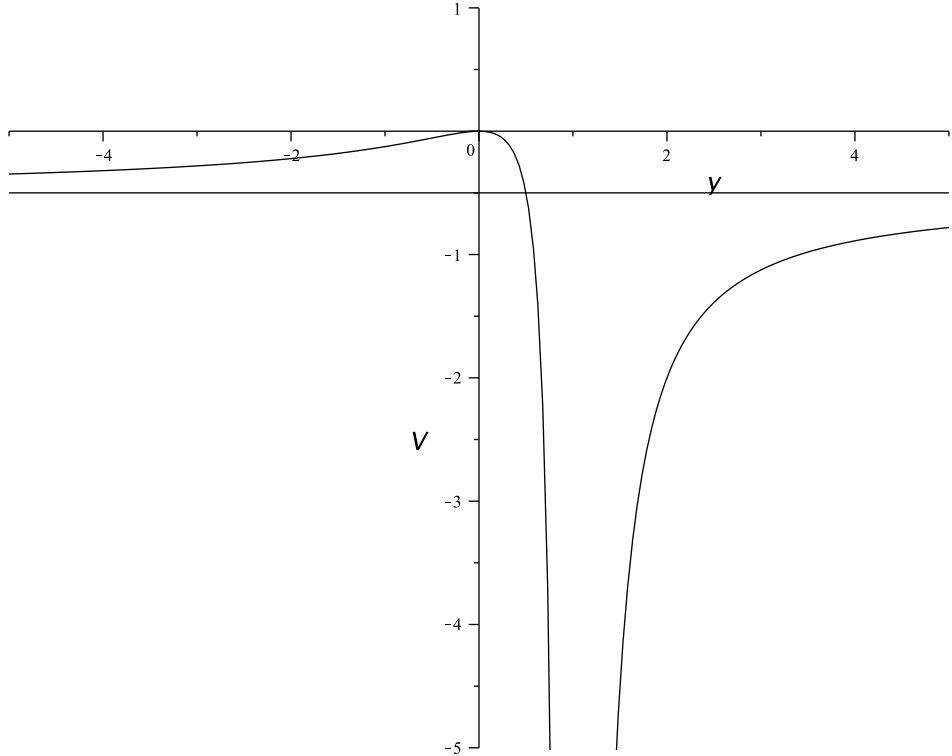}
\caption{The potential $V(y)$ for $\lambda=1$. The 
horizontal line $E=-1/2$ intersects the
graph of $V(y)$ only once for $y>0$ and there is a region, 
bounded by this intersection and by the vertical asymptote 
$y=C/\lambda$, which describes bounded 
motion. (The parameter value $C=1$ is used in this  plot.)  
\label{fig:lambda1}} \end{figure}

If $E=-\frac{1}{2}=-\frac{1}{2\lambda^2}$, corresponding to 
$\lambda=1$,  there is only unbounded 
motion in the region $y>C/\lambda$, but there is a region 
of bounded motion
\be
y_3=\frac{C}{2} \leq y(x) < \frac{C}{\lambda} 
\ee
which can make sense as the transverse profile of a 
glacial 
valley (see fig.~\ref{fig:lambda1}).

As $\lambda \rightarrow 1^{+}$, one of the turning 
points~(\ref{y12}) is pushed to infinity and effectively 
disappears, leaving a single turning point $y_3=C/2$.

\subsection{The case $0 <\lambda<1$}

If $ -\frac{1}{2\lambda^2}< E=-\frac{1}{2}$, corresponding 
to $\lambda<1$, the horizontal line of constant energy 
$E=-1/2$ lies above the horizontal asymptote of $V(y)$ and 
intersects the graph of $V(y)$ only once in the region 
$y>0$. There is a region of bounded motion 
\be
y_4 \leq y(x) < \frac{C}{\lambda} \,.
\ee
The situation is qualitatively similar to the $\lambda=1$ 
case. The turning point $y_4=C/(1-\lambda) $ lies in the 
$y>0$ region, while the second turning point $y_5 
=-C/(1+\lambda)$ lies in the uninteresting region 
$y<0$. The analytic solution~(\ref{Gibbonssolution}) of 
eq.~(\ref{7}) found in 
Ref.~\cite{Gibbons0} belongs to this situation.

\subsection{The case $\lambda <0$}

\begin{figure}
\centering
\includegraphics[scale=0.65]{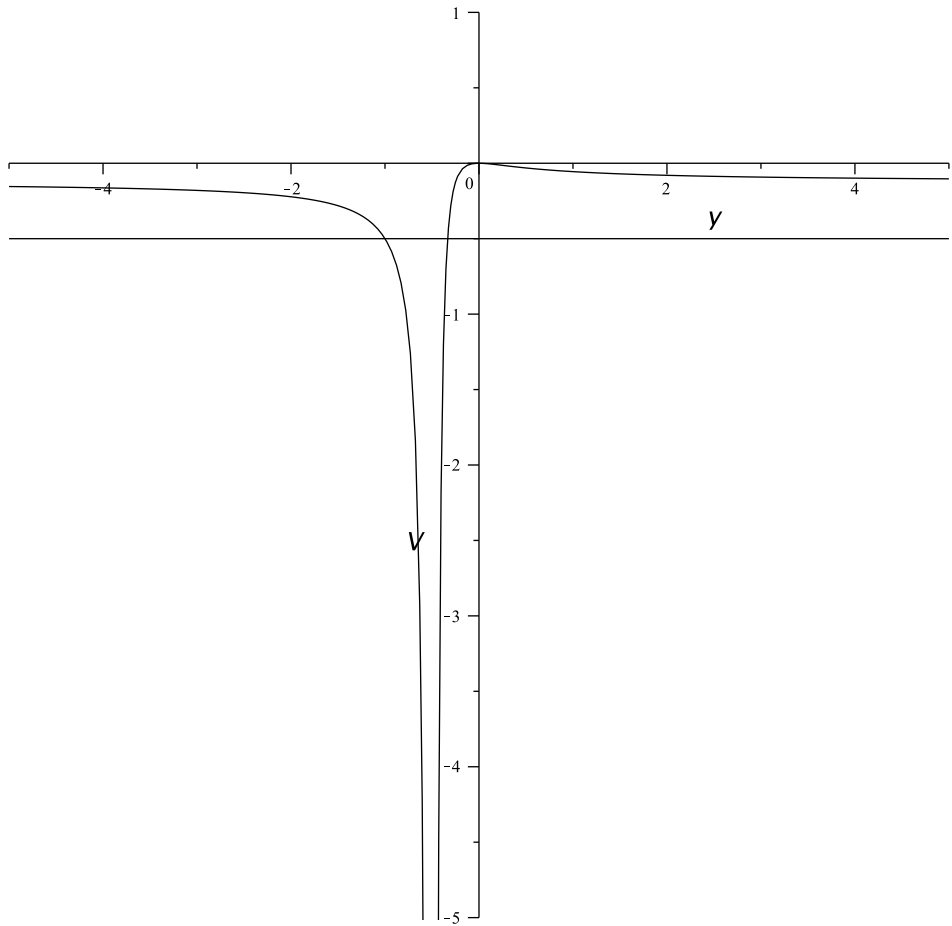}
\caption{The potential $V(y)$ for negative $C$. 
In this case the motion is 
forbidden in the $y>0$ region if $ E=-\frac{1}{2} \leq 
-\frac{1}{2\lambda^2}$. 
If $ -1< \lambda <0 $ the line $E=-1/2$ intersects the 
graph of $V(y)$ only at $y_6= C/(1+\lambda) $ in the $y>0$ 
region 
and there are no bounded motions. (The 
parameter values $\lambda=2$, 
$C=-1$ are  chosen for illustration.)
\label{fig:negC}}
\end{figure}

We can now comment on the second condition $C>0$ appearing 
in Ref.~\cite{Morgan05} and assumed at the beginning of 
this section. Given the symmetry of eq.~(\ref{7}),  the 
situation $C<0$ and $\lambda>0$ is 
equivalent to  $C>0$ and $\lambda<0$, which we discuss 
here. In this case  the vertical 
asymptote $y=C/\lambda$ of $V(y)$ lies in the $y<0$ 
region, and $V'(y)$ is negative 
for $y<C/\lambda<0$ and for $y>0$, and is positive for 
$C/\lambda <y<0$. The graph of the potential energy $V(y)$ 
in this case is shown in fig.~\ref{fig:negC}. 
If $E=- \frac{1}{2} > -\frac{1}{2\lambda^2}$ (corresponding 
to $ -1< \lambda <0 $), there is only one intersection 
$y_6= C/(1+\lambda) $ in the $y>0$ region and there are no 
bounded motions.

In the remaining case $C=0$, eq.~(\ref{7}) reduces to 
$(y')^2= -1 +1/\lambda^2$, which has linear solutions 
corresponding to V-shaped valleys irrelevant as glacial 
valley cross-profiles (except perhaps as initial 
conditions, see, {\em e.g.}, \cite{numerical2}). This is 
the reason why we assumed that $C>0$ and we will restrict 
to this range of the parameter $C$ in the rest of this 
work.

\subsection{Analogue of the no-constraint equation}

Finally, we comment on the incorrect equation~(\ref{1.17}) 
which would be obtained by extremizing friction without any 
Lagrangian constraint. Using the mechanical analogy, 
eq.~(\ref{1.17}) can be rewritten as the energy integral of 
motion
\be
\frac{(\xi')^2}{2} +U( \xi) =E \,, 
\ee
where $E=-1/2$ and the potential energy
\be
U( \xi )= -\frac{\xi^2}{ 2C_2^2 } 
\ee
corresponds to an inverted harmonic oscillator. 
Therefore, all trajectories (except for the unstable 
equilibrium position $\xi \equiv 0$, which is meaningless 
in the original valley glacier problem) are unbounded and 
unphysical as glacier cross-profiles.

\section{The universe in a glacial valley}
\setcounter{equation}{0}
\label{sec:3}

Relativistic cosmology ({\em e.g.}, \cite{Wald, Carroll, 
Liddle}) is 
obtained by assuming that the 
4-dimensional spacetime of general relativity is spatially homogeneous and 
isotropic about every point of 3-space. This assumption is 
motivated by the fact that, on scales of hundreds of  
megaparsecs,  the matter distribution and the cosmic 
microwave background are spatially homogenous and 
isotropic \cite{Liddle, Wald, Carroll}. 
The cosmic microwave background, in particular, 
is highly isotropic apart from tiny temperature 
fluctuations $\delta T/T \sim 10^{-5}$ 
imprinted early on by large-scale structures, which play a major 
role in our investigations of the early universe \cite{Durrer}. 
These assumptions lead uniquely to the 
Friedmann-Lema\^itre-Robertson-Walker (FLRW) line 
element of spacetime \cite{Wald, Carroll, Liddle}
\be\label{FLRW}
ds^2 =-dt^2 +a^2(t) \left[ \frac{dr^2}{1-k r^2} +r^2 \left( 
d\theta^2 + \sin^2 \theta \, d\varphi^2 \right) \right] 
\ee
in polar coordinates $\left( t,r, \theta, \varphi \right)$, 
where $t$ is the time of the observers who see the cosmic 
microwave background homogeneous and isotropic around them 
apart from the tiny fluctuations ({\em comoving 
observers}). The positive 
function $a(t)$ ({\em scale factor}) describes how any two 
points of space (for example two typical comoving galaxies)
separate in time as the universe expands (which is 
described 
by an increasing function $a(t)$). $k$ is the {\em 
curvature index} normalized to the possible values $k=-1$ 
(open universe), $k=0$ (critically open or spatially flat 
universe), or $k=+1$ (closed universe) \cite{Wald, 
Carroll, Liddle}. Following the literature on cosmology, 
we use units in which the speed of light $c$ is unity. 
It is usually 
assumed in cosmological investigations that the material  
content of the universe is in the form of a perfect fluid 
of energy density $\rho$ and pressure $P$, as measured by 
the comoving observers. Once an equation of state for the 
fluid linking $P$ and $\rho$ is specified, one can solve 
the Einstein equations for the spacetime metric $g_{ab}$ 
giving the line element through $ds^2= g_{ab}dx^a dx^b$ 
(we use the Einstein 
summation convention over repeated indices). 
Because, due to spatial homogeneity 
and isotropy, the line element must assume the FLRW form 
(\ref{FLRW}), 
the only degree of freedom is the scale factor $a(t)$ and 
the Einstein field equations, which are usually PDEs, 
reduce to the Einstein-Friedmann ODEs for $a(t)$ 
\cite{Wald, Carroll, Liddle}
\begin{eqnarray}
H^2 &\equiv & \frac{\dot{a}^2}{a^2}= \frac{8\pi G}{3} \, 
\rho -\frac{k}{a^2} \,,\label{Hconstraint}\\
&&\nonumber\\
\frac{ \ddot{a}}{a} &=& -\frac{4\pi G}{3} \left( \rho +3P 
\right) \,,\label{acceleration} 
\end{eqnarray}
where an overdot denotes differentiation with respect to 
the comoving time $t$ and $H (t) \equiv \dot{a}/a$ is the 
Hubble function. $G$ is Newton's constant expressing the 
strength of the coupling between gravity and matter. 
Eq.~(\ref{Hconstraint}) 
is often referred to as the {\em Friedmann equation}.

In general relativity, the pressure $P$ of the 
cosmic fluid gravitates together with 
the energy 
density $\rho$ and the combination $\rho+3P$ in 
eq.~(\ref{acceleration}) determines whether the universe 
accelerates ($\ddot{a}>0$) or decelerates ($\ddot{a}<0$) 
its expansion.

A third convenient (but not independent)  equation can be 
derived from the previous two:
\be\label{conservation}
\dot{\rho}+3H \left( P+\rho \right)=0 
\ee
and it expresses covariant conservation of the 
energy-momentum tensor of the cosmic fluid \cite{Wald, 
Carroll, Liddle}.

The Friedmann equation~(\ref{Hconstraint}) is the analogue 
in cosmology of 
eq.~(\ref{7}) describing the ice thickness in glacial 
valley transverse 
profiles. The analogy was noted recently, but not pursued, 
in Ref.~\cite{Gibbons}. Although one can attempt an analogy 
with universes corresponding to different values of the 
curvature index $k$ in eq.~(\ref{FLRW}), the most 
straightforward identification between eqs.~(\ref{7}) and 
eq.~(\ref{Hconstraint}) is achieved by setting $k=1$, which 
corresponds to a closed universe analogue of the glacial 
valley profile. The cosmological analogue of eq.~(\ref{7})
\be \label{3.5}
\frac{ \dot{a}^2}{a^2} =\frac{1}{\left( \lambda a- C 
\right)^2} -\frac{1}{a^2}
\ee
can be rewritten as 
\be \label{Hconstraintt}
\frac{ \dot{a}^2}{a^2} =\frac{8\pi G}{3} \, 
\frac{\rho_0}{\left( a- a_0 \right)^2} -\frac{1}{a^2} \,,
\ee
where
\be \label{qqq}
\rho_0 = \frac{3}{8\pi G \lambda^2}  \,, \;\;\;\;\;
a_0 = \frac{C}{\lambda}
\ee
are positive constants. 
Eq.~(\ref{Hconstraint}) gives immediately the energy 
density of the analogue cosmic fluid as
\be \label{density}
\rho(t) = \frac{ \rho_0}{\left( a-a_0 \right)^2} \,.
\ee
Two properties of the function $\rho(t)$ are relevant. 
First, the energy density is 
always positive, which is expected of ``reasonable'' forms 
of  matter but is not guaranteed in any formal analogy of 
ODEs with 
the equations of FLRW cosmology (indeed, the 
identification of eq.~(\ref{7}) with the  
cosmological equation~(\ref{Hconstraint}) with 
curvature index $k \leq 0$ 
gives rise to negative effective energy densities $\rho$ 
when $a \rightarrow 0$, an unphysical property which would 
detract from the 
analogy). Second, the density 
diverges if $a \rightarrow a_0$: this divergence 
corresponds to a true 
spacetime singularity (as opposed to a coordinate 
singularity) and is discussed below.

``Reasonable'' matter in general relativity is supposed to 
satisfy certain energy conditions which, essentially, 
prohibit negative energy densities and energy flows faster 
than the speed of light \cite{Wald, Carroll, Liddle}. When 
applied 
to a perfect fluid with energy density $\rho$ and pressure 
$P$, the two energy conditions which are most often 
encountered in the literature (and that are relevant 
below) are the {\em weak energy 
condition}, which amounts to $\rho \geq 0$ and $\rho+P \geq 
0$, and the {\em strong energy condition} $\rho+P \geq 0$ 
and $\rho+3P \geq 0$ \cite{Carroll, Wald, Liddle}.

Let us proceed by deducing the effective pressure $P$ of 
the analogue cosmic fluid by imposing 
eq.~(\ref{conservation}), which 
yields 
\be\label{pressure} 
P=\frac{2\rho_0 a}{3\left( a-a_0 
\right)^3} -\frac{\rho_0}{\left( a-a_0 \right)^2} 
\ee 
and can be rewritten as 
\be
\label{eos} 
P = -\frac{\rho}{3} 
\pm\frac{2a_0}{3\sqrt{\rho_0}}\, \rho^{3/2} \,, 
\ee 
where 
the upper sign applies when $ a>a_0$ and the lower sign 
when $ a<a_0$. Using eq.~(\ref{pressure}), it is seen that 
the 
condition $\rho+ P >0$ corresponds to 
$a>a_0$ and, {\em vice-versa}, violation of this condition 
corresponds to $a<a_0$. Equations of state of the 
cosmic fluid corresponding to the lower sign in 
eq.~(\ref{eos}) and violating the weak energy condition  
have been discussed in 
Ref.~\cite{BarrowSFS}. Equations of state of the cosmic 
fluid of the form $P=\sum_{k=1}^m c_k \rho_{(k)}^k$ have 
been studied in \cite{Gibbons0, Gibbons}. Quadratic 
equations of state, in particular, have been the subject of 
further attention \cite{quadratic, quadratic2, quadratic3, 
quadratic4, quadratic5, quadratic6} but pressures depending 
on fractional powers of the density have not been 
studied. As shown below, they give rise to a 
peculiar 
type of singularities. While traditional cosmology and 
relativity textbooks report only linear barotropic 
equations of state $P=P_0+A\rho$, following the discovery 
of the acceleration of the universe in 1998, the literature 
abounds with exotic non-linear equations of state for the 
dark energy fluid postulated in order to explain this 
acceleration.

Let us consider now the acceleration 
equation~(\ref{acceleration}) which, using 
eq.~(\ref{pressure}), reduces to 
\be
\frac{ \ddot{a}}{a} = - \frac{a_0}{\lambda^2 \left(
a-a_0\right)^3} \,.
\ee
Clearly, the universe is accelerated if $a <a_0$ 
(corresponding to $\rho+3P<0$) and 
decelerated if  $a>a_0$ (corresponding to $\rho+3P >0$). 

The value $a_0$ of the scale factor corresponds to a 
spacetime singularity, which is seen as follows. The 
Einstein equations
\be
R_{ab}-\frac{1}{2} \, g_{ab} R =8\pi G T_{ab} 
\ee
(where $R_{ab}$ is the Ricci tensor and $R \equiv 
g^{ab}R_{ab}$ is its trace, while $g^{ab}$ is the inverse 
of the metric tensor $g_{ab}$) can be traced to give
\be
R \equiv {R^a}_a =-8\pi G T \,,
\ee
where $T \equiv {T^a}_a $ is the trace of the perfect fluid 
energy 
momentum 
tensor \cite{Wald, Carroll, Liddle}
\be\label{perfectfluid}
T_{ab}=\left( P+\rho \right) u_a u_b +P g_{ab}
\ee
and $u^a$ is the 4-velocity of the cosmic fluid 
(equivalently, the 4-velocity of comoving observers).  One 
obtains the Ricci scalar
\be R=8\pi G \left( \rho -3P \right) = \frac{16\pi G 
\rho_0}{\left( a-a_0 \right)^3} \left( a-2a_0 \right) \,,
\ee
which diverges in the limit $a \rightarrow a_0$, signalling 
a spacetime singularity.  However, it is not yet clear 
whether the value $a_0$ of the scale factor, which 
corresponds to a spacetime singularity and diverging 
density and pressure, can actually be approached during the 
dynamics of the scale factor $a(t)$. In order to 
answer this question note that eq.~(\ref{Hconstraint}), 
which is of only first order, constitutes a dynamical 
constraint\footnote{In general, the dynamics of general 
relativity is constrained dynamics and FLRW cosmology is no 
exception \cite{Wald, Carroll, Liddle}.} since it must 
be $H^2 \geq 0$, which can be written as 
\be\label{constraint}
a^2 \geq \lambda^2 \left( a-a_0 \right)^2 
\ee
using $k=+1$. 
The condition~(\ref{constraint}) excludes the orbits of the 
solutions of the 
dynamical system~(\ref{Hconstraint}), (\ref{acceleration}) 
from a certain volume of the $\left( a, \dot{a} \right)$ 
phase space. This constraint plays the role that the first 
integral expressing conservation of energy plays in point 
particle mechanics by confining the orbits of the solutions 
to an energy surface $E=$~const. in phase space.

If $a>a_0$, the dynamical constraint~(\ref{constraint}) can 
be written as $ a \leq \lambda a_0 /( \lambda -1)$. The 
coefficient of $a_0$ is   
$\frac{ \lambda}{\lambda -1}  
=1+\frac{1}{\lambda-1} >1$ and therefore, in this regime we 
have
\be \label{3.16}
a_0 < a < \frac{ \lambda a_0}{ \lambda -1} \,:
\ee
the scale factor $a(t)$ cannot grow arbitrarily large but 
is bounded from above. However, it can get arbitrarily 
close to the value $a_0$ corresponding to the 
singularity.

If instead $a< a_0$, then the constraint~(\ref{constraint}) 
becomes $a \geq \lambda a_0 /(\lambda+1)$, hence in this 
regime one has
\be\label{3.17}
0< \frac{\lambda a_0}{\lambda +1} \leq a < a_0 \,.
\ee
In both cases the scale factor is bounded from below by a 
positive constant  
(hence one cannot have a Big Bang- or Big Crunch-type 
singularity, which 
would correspond to $a \rightarrow 0$ \cite{Wald, 
Carroll, Liddle}) but it can reach the singularity 
$a_0$. The fact that one cannot have $y=0$ in the original 
equation~(\ref{7}) because it implies an imaginary $y'$ 
was noted as  ``a curious feature'' in 
Ref.~\cite{Morgan05}. It corresponds to the fact that $y=0$ 
(equivalently, $a=0$) lies in the region of the phase space 
forbidden by the constraint~(\ref{constraint}).

The boundary values $a=\lambda a_0/(\lambda \mp 1)$ 
are formal solutions of eq.~(\ref{3.5}) obtained by setting 
$a=$~constant, but they do not satisfy 
eq.~(\ref{acceleration}) (they would be meaningless  
anyway as analogues of glacial valley profiles).

\begin{figure}
\centering
\includegraphics[scale=0.65]{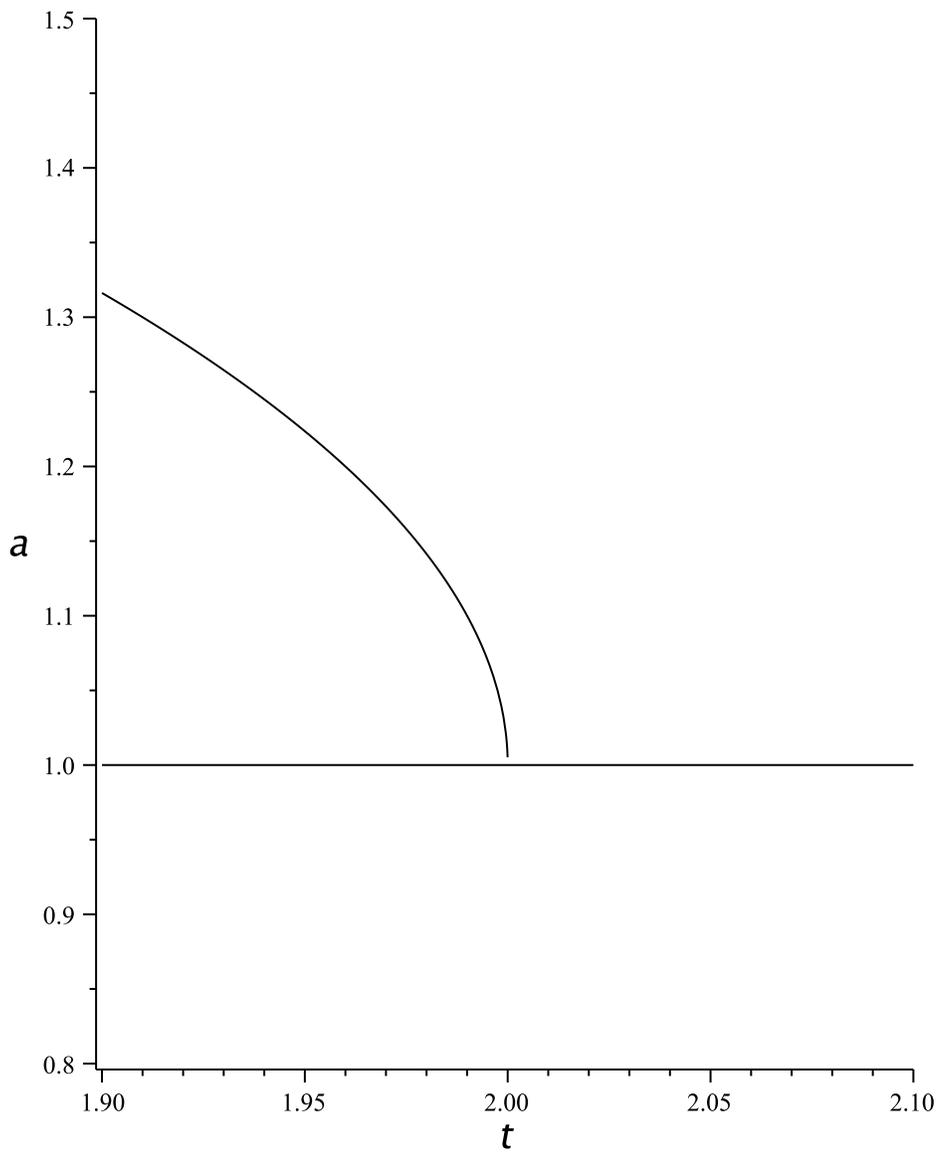}
\caption{The behavior of the scale factor $a(t)$ near the 
singularity $a_0$ when $a>a_0$ (for the parameter values 
$a_0=1$, $t_0=2$). The concavity faces downward and the 
universe is decelerated. 
\label{fig:3}}
\end{figure}

Consider again the acceleration 
equation~(\ref{acceleration}): 
if $a > a_0 $, then $\ddot{a}<0$ and the curve 
representing the scale factor has concavity facing 
downwards, describing a decelerated universe. Since this 
curve is continuous, 
it must always decrease and eventually cross the horizontal 
line $a=a_0$ (see fig.~\ref{fig:3}). 
Using eq.~(\ref{qqq}), eq.~(\ref{3.5}) is written as 
\be \label{3.5rewritten}
\dot{a}^2= \frac{a^2}{\lambda^2 \left( a-a_0 \right)^2} 
-1 \approx \frac{a_0^2}{\lambda^2 \left( a-a_0 \right)^2}
\ee
as $ a\rightarrow a_0$. This asymptotic equation is easily 
integrated, giving
\be
a(t) \simeq a_0 \pm \sqrt{ \frac{2a_0}{\lambda} | t-t_0 |} 
\,,
\ee
where the integration constant $t_0$ has the meaning of 
time at which the singularity occurs and the positive sign 
must 
be chosen in front of the square root because it is 
$a>a_0$. This situation  
constitutes a physically meaningful analogue of glacial 
valleys because $y$ is the {\em thickness} of the ice 
(maximum at $x=0$ and minimum at the valley boundaries) and 
it is interesting in cosmology because it 
provides an example of a finite time singularity even when 
$\rho >0$ and $\rho+P>0$, {\em i.e.}, without violating the 
weak energy condition. This kind of situation was discussed 
in Ref.~\cite{BarrowSFS}. 

\begin{figure}
\centering
\includegraphics[scale=0.65]{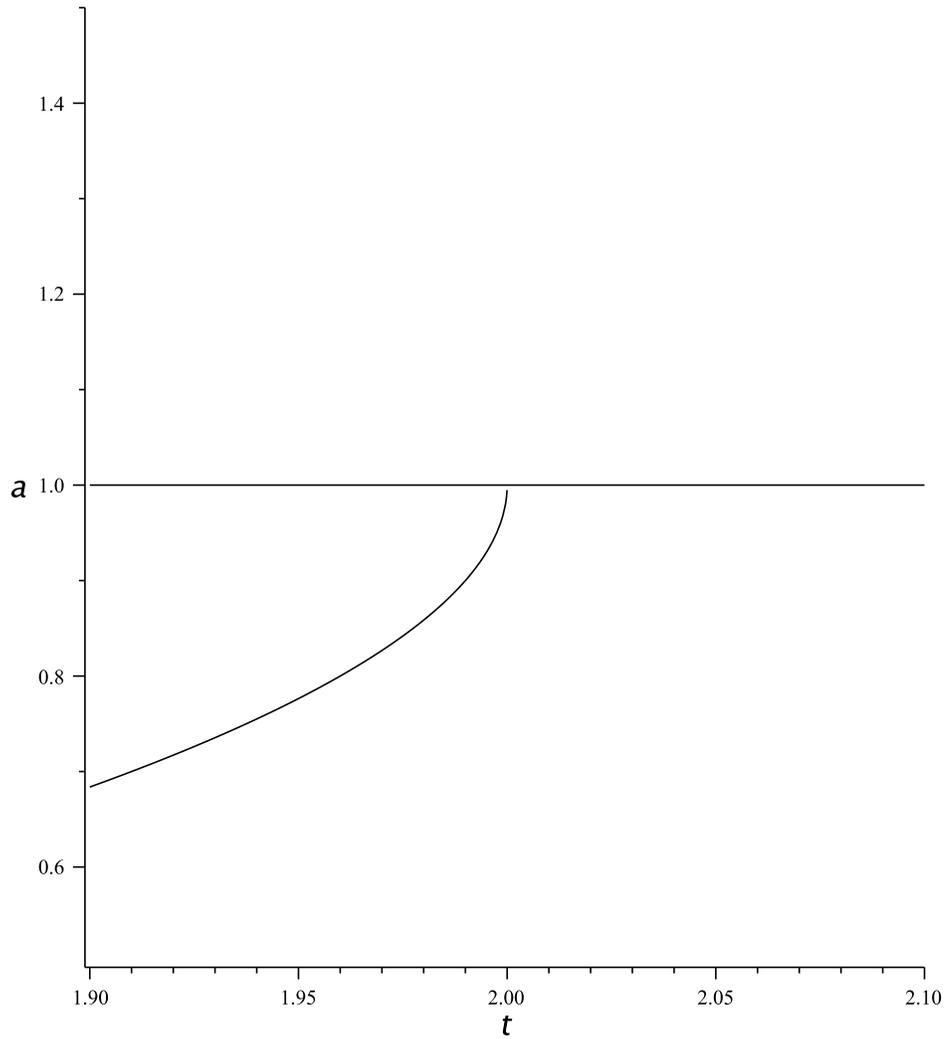}
\caption{The behavior of the scale factor $a(t)$ near the 
singularity $a_0$ when $a<a_0$ (for the parameter values 
$a_0=1$, $t_0=2$). The concavity faces upward and the 
universe is accelerated.
\label{fig:4}}
\end{figure}

{\em Vice-versa}, if $a<a_0$, then it is $\ddot{a}>0$ and  
$a(t)$ always increases, eventually crossing the horizontal 
line 
$a=a_0$ (fig.~\ref{fig:4}). 
The asymptotic equation~(\ref{3.5rewritten}) is now 
integrated to 
\be
a(t) \simeq a_0-\sqrt{ \frac{2a_0}{\lambda} |t-t_0|}
\ee
choosing the negative sign in front of the square root 
because it is now $a<a_0$. This situation is not a 
meaningful analogue 
of a glacial valley cross-profile. The slope of this 
function becomes infinite where 
$a(t)$ crosses the value $a_0$, and this boundary 
corresponds to the spacetime singularity in the 
cosmological analogue. In this case the universe has a 
minimum size $a_{min}=\lambda a_0/ (\lambda +1)$ 
and it bounces upon reaching it. It is well 
known in cosmology that such a bounce occurs when  
the weak energy condition $\rho+P<0$ is violated, which 
is exactly what is happening here since
\be
\rho +P = \frac{2\rho_0 a}{3\left( a- a_0 \right)^3} <0 
\,.
\ee
The violation of the weak energy  condition signals that 
the cosmic fluid is of a 
very exotic form referred to as {\em phantom energy}, which 
causes the universe to accelerate in such a way that the 
Hubble function increases according to 
\be
\dot{H}=-4\pi G \left( \rho +P \right) + \frac{k}{2a^2}>0 
\ee
if $k \geq 0$. (This equation can be obtained by 
differentiating eq.~(\ref{Hconstraint}) and substituting 
eqs.~(\ref{Hconstraint}) and (\ref{acceleration}) into the 
result.) This is a very unusual situation (called {\em 
superacceleration}) and causes the universe to expand 
super-exponentially and reach a singularity at a finite 
time. In the standard cosmological literature the scale 
factor of a phantom-dominated universe 
diverges at a finite time in the future at a Big Rip 
singularity \cite{Caldwell}, but here the situation is 
different since the scale factor stays finite while the 
Hubble function $H$, energy density $\rho$, pressure $P$, 
and Ricci scalar $R$ all diverge as $ a\rightarrow a_0$.  
This situation corresponds instead to a type of singularity 
studied recently in spatially flat ({\em i.e.}, 
$k=0$) universes and called Big 
Freeze singularity \cite{Mariam} or Type~III singularity 
in 
the classification of \cite{Bamba}. A Big Freeze 
singularity appears also in cosmology in the context of 
Palatini $f(R)$ gravity \cite{B1, B2, B3, B4}. This 
is a class of 
theories of 
gravity alternative to general relativity and attempting to 
explain the present acceleration of the universe without 
dark energy.  A Big Freeze  
singularity was not reported before for positively curved 
({\em i.e.}, $k>0$) universes.  The reason why here the 
situation is essentially 
the same as for spatially flat universes is that, as 
$a\rightarrow a_0$ in eq.~(\ref{Hconstraintt}), the 
divergent 
term proportional to $\left(a-a_0 \right)^{-2}$ in the 
energy density dominates over the curvature term $- 1/a^2$, 
which is finite and becomes irrelevant. Finite time 
singularities, including Big Rip 
\cite{Caldwell} and sudden future singularitities 
\cite{BarrowGallowayTipler, BarrowSFS}, have been the 
subject of a significant amount of work in cosmology (see 
\cite{Calcagni, Gorini, Kofinas, Sahni, Dabrowski2, 
Dabrowski, Fernandez, Mariam, Bamba, Diego} and the 
extensive literature originating from these references).

Finally, although it is not interesting for the original 
glacial valley problem, the cosmic analogue of 
eq.~(\ref{7}) for $C=0$ (noted also in \cite{Gibbons}) 
corresponds to $a_0=0$ and 
\be
H^2=\frac{1}{a^2} \left( 
\frac{1-\lambda^2}{\lambda^2}\right)
\ee
and requires $|\lambda|\leq 1$. Then 
$\dot{a}=$~const. and the solutions are linear in 
time and include a static universe with $a=$~const. as a 
special case.

\section{Discussion}
\setcounter{equation}{0}
\label{sec:4}

In glacial morphology studies, researchers content 
themselves with fitting data of glacial valley 
cross-profiles with parabolas $y(x) =ax^2 +bx +c $ 
(following an early practice initiated by Svennsson 
\cite{Svensson} which is not free of critique, see {\em 
e.g.}, \cite{PattynHuele}). Other fitting curves used 
include power-law profiles $y(x)=a x^b$, possibly with 
different powers $b$ for each half-profile going 
from the bottom at $x=0$ to each valley side.  

There is a deep disconnect between theory and practice  
here. Since a parabola is just the second order Taylor 
expansion 
of {\em any} sufficiently regular function with a minimum, 
which could   
solve any ODE, data fitting with parabolas is of no 
help when one attempts to test models and to discriminate 
between various theoretical approaches to the problem 
of the cross-sectional profiles of glacial  
valleys, which predict different ODEs for the profile 
$y(x)$. In this sense fitting parabolas, or 
even power-laws, is deeply unsatisfactory from the 
theoretical point of view. Determining the best-fit 
parameters of a parabola for a glacial valley, or an ensemble 
of valleys, does not contribute to understanding the 
mechanism that generated them.

Here we focused on the variational principle approach to 
the problem of glacial valley erosion, in the form given by 
Morgan \cite{Morgan05} which imposes that the 
cross-sectional area of the valley is fixed as a Lagrangian 
constraint. This condition is also used in modern numerical 
modelling of the detailed glacial erosion process 
({\em e.g.}, \cite{numerical2}). The 
resulting eq.~(\ref{7}) for the ice thickness 
 $y(x)$ has 
an analogue in point particle 
mechanics and one in cosmology, both of which contribute 
to a better understanding of the character of this ODE, 
of its solutions, and of the conditions (on the 
parameters $\lambda$ and $C$) for their 
physical viability. We developed these analogies in 
detail. The previous analysis \cite{Morgan05} 
is augmented by the graphical analysis of the 
effective potential $V(y)$ in the mechanical 
analogy. In turn, the problem of glacial valley 
profiles provides an interesting example of a finite time 
singularity in cosmology without violating the weak 
energy condition. The finite time singularity is caused by 
the peculiar effective equation of state~(\ref{eos}) which falls 
into the broader category $P=w\rho -\alpha \rho^m$ (with 
$w, \alpha$, and $m$ constants) which has been the subject of 
wide interest in cosmology but restricted 
to integral values of the exponent $m$ \cite{BarrowSFS, 
Calcagni, Gorini,  Kofinas, Sahni}.

The present work also highlights two open problems in 
glaciology. First, the friction of glacier ice against the 
valley walls and bed is unlikely to be described purely by 
Coulomb's law, but it should include viscous friction which 
depends on the velocity (a friction model quadratic in the 
velocity is used, for example, in the numerical work  
\cite{numerical2}). Second, the numerical analyses of the 
formation of glacial valleys ignore the variational 
approach (but not its Lagrangian constraint) and should be 
compared with it. These issues will be revisited in future 
work.

\section*{Acknowledgments} This work is supported by  the 
Natural Sciences and Engineering Research Council of 
Canada (NSERC).

\appendix
\section*{Appendices}
\renewcommand{\thesubsection}{\Alph{subsection}}

\subsection{The catenary problem}
\label{appendix:A}
\def\theequation{A.\arabic{equation}}\setcounter{equation}{0}

The classic catenary problem ({\em e.g.}, \cite{Goldstein}) 
gives rise to the same ODE~(\ref{6}) obtained by Hirano and 
Aniya in \cite{HiranoAniya88}. Consider a heavy string 
hanging in a vertical $\left( x, y \right)$ plane and 
described by the profile $y(x)$. The linear density is 
$\mu=dm/ds$, where $ds=\sqrt{dx^2+dy^2}=\sqrt{ 1+(y')^2}\, 
dx $ is the line element along the string.  The 
gravitational potential energy of an element of string of 
length $ds$ located at horizontal position $x$ is $dE_g= 
\mu gy(x) ds $. The total gravitational potential energy of 
a string suspended by two points of horizontal coordinates 
$x_1$ and $x_2$ is the functional of the curve $y(x)$
\be
E_g \left[ y(x) \right] = \mu g \int_{x_1}^{x_2} dx \, y 
\sqrt{1+(y')^2} \equiv \int_{x_1}^{x_2} L \, dx 
\,.
\ee
The Lagrangian $L\left( y(x), y'(x) \right) $ does not 
depend explicitly on the coordinate $x$ and, therefore, the 
corresponding Hamiltonian is conserved:
\be
{\cal H} =\frac{\partial L}{\partial (y')} \, y'-L=c_1 \,,
\ee 
where $c_1$ is a constant. This equation simplifies to 
\be \label{appendix-questa}
\frac{-y}{\sqrt{ 1+(y')^2}}=c_1 \,,
\ee
which has catenaries as solutions \cite{Goldstein}. 
With the change of variable $y \rightarrow \bar{y} \equiv 
y-y_s -\lambda$, eq.~(\ref{6}) reduces to the same form as 
eq.~(\ref{appendix-questa}) and it is not surprising 
that Hirano and 
Aniya find catenary solutions \cite{HiranoAniya88}.

Although not necessary, the variational principle is 
sometimes imposed subject to the constraint that the  
string length between $x_1$ and $x_2$ is fixed, 
which changes the variational integral to 
\be
J \left[ y(x) \right] = \int_{x_1}^{x_2} dx \, \left( y+\lambda \right) 
\sqrt{1+(y')^2} \,,
\ee
where $\lambda $ is a Lagrange multiplier.\\\\

The cosmological analogue of eq.~(\ref{appendix-questa}) 
corresponds to a well known situation in cosmic physics. This equation is 
recast as 
\be
\left( \frac{ y'}{y} \right)^2 = \frac{1}{C_1^2}-\frac{1}{y^2} \,,
\ee
the analogue of which for the scale factor $a(t)$ reads
\be
\frac{ \dot{a}^2}{a^2} = 
\frac{\Lambda}{3}-\frac{1}{a^2} \,,
\ee
where $\Lambda= 3/C_1^2 >0$ is the cosmological constant. 
The cosmological constant  corresponds to the modification 
of the Einstein equations \cite{Wald, Carroll, Liddle}
\be
R_{ab}-\frac{1}{2} \, g_{ab} R +\Lambda g_{ab}=8\pi G 
T_{ab} \,,
\ee
but it can be treated as a special case of a perfect fluid 
by taking the $\Lambda$-term to the right hand side and 
treating it as an effective stress-energy tensor 
$ T_{ab}=- \Lambda g_{ab}/(8\pi G) $, which has the perfect fluid 
form~(\ref{perfectfluid}) with energy density and pressure 
\cite{Wald, Carroll, Liddle}
\be
\rho_{\Lambda}= \frac{\Lambda}{8\pi G}=-P_{\Lambda} \,.
\ee
A positive cosmological constant violates the strong energy 
condition $\rho +3P \geq 0$ and essentially corresponds
to repulsive gravity \cite{Carroll, Liddle, 
AmendolaTsujikawabook}. The de Sitter 
expansion $a(t) = \mbox{const.} \, \mbox{e}^{Ht}$ with 
constant Hubble function $H$ gives a solution of the 
Einstein-Friedmann 
equations with $\Lambda$ as the only material source if 
$k=0$ 
and is an attractor for the $k=0$ spaces filled with 
other matter sources in addition to $\Lambda$, and 
also for the $k\neq 0$ cases.  The importance of 
the de Sitter space in cosmology can hardly be 
overemphasized, since this solution is an attractor for the 
regime of slow-roll inflation in the early universe 
\cite{KolbTurner, Liddle} and in the late universe 
\cite{AmendolaTsujikawabook}.

\end{document}